# Atomistic Mechanisms of Mg Insertion Reactions in Group XIV Anodes for Mg-Ion Batteries

Mingchao Wang[*,†], Jodie A. Yuwono[†], Vallabh Vasudevan[†], Nick Birbilis[†], Nikhil V. Medhekar[*,†]

[†]Department of Materials Science and Engineering, Monash University, Clayton, VIC 3800, Australia

[*]Email: mingchao.wang@monash.edu (Mingchao Wang); nikhil.medhekar@monash.edu (Nikhil V. Medhekar).

**Abstract:** Magnesium (Mg) metal has been widely explored as an anode material for Mg-ion batteries (MIBs) owing to its large specific capacity and dendrite-free operation. However critical challenges, such as the formation of passivation layers during battery operation and anode-electrolyte-cathode incompatibilities, limit the practical application of Mg-metal anodes for MIBs. Motivated by the promise of group XIV elements (namely Si, Ge and Sn) as anodes for lithium- and sodium-ion batteries, here we conduct systematic first principles calculations to explore the thermodynamics and kinetics of group XIV anodes for Mg-ion batteries, and to identify the atomistic mechanisms of the electrochemical insertion reactions of Mg ions. We confirm the formation of amorphous $Mg_xX$ phases (where X = Si, Ge, Sn) in anodes via the breaking of the stronger X-X bonding network replaced by weaker Mg-X bonding. Mg ions have higher diffusivities in Ge and Sn anodes than in Si, resulting from weaker Ge-Ge and Sn-Sn bonding networks. In addition, we identify thermodynamic instabilities of $Mg_xX$ that require a small overpotential to avoid aggregation (plating) of Mg at anode/electrolyte interfaces. Such comprehensive first principles calculations demonstrate that amorphous Ge and crystalline Sn can be potentially effective anodes for practical applications in Mg-ion batteries.

**Keywords:** *Mg-ion batteries*, *group XIV elements*, *anode materials*, *first-principles calculations*, *ionic diffusivity*, *phase transition*, *overpotential*, *electrode potential*.

## 1. INTRODUCTION

Rechargeable Mg-ion batteries (MIBs) are rapidly emerging as a promising technology for energy storage and conversion, owing to several features that are potentially advantageous.[1-2] The utilization of divalent Mg as a charge-carrying ion offers the prospect of developing high capacity, low cost and safe battery systems relative to existing battery systems. To realize such beneficial features of MIBs, various electrode and electrolyte materials are presently being explored.[3] For instance, a large number of intercalation cathodes, such as $MnO_2$ polymorphs,[4-5] Chevrel $Mo_6S_8$,[6-7] and layered chalcogenides $MX_2$ (M = Mo, Ti; X = S)[8-9] have been confirmed as high energy-density cathode materials for MIBs. In these studies, Mg metal is typically selected as the anodic counterpart due to its high specific volumetric capacity of 3833 $mAh/cm^3$, along with Mg offering safe operation owing to the absence of dendrites, in comparison with Li metal in Li-ion batteries.[10-11]

The unique and complex electrochemistry of Mg has to date limited the selection of electrolytes that satisfy the requirements of compatibility with high-voltage cathodes while ensuring a reversible plating (deposition)/stripping (dissolution) reactions of the Mg anode during normal battery operation. To fulfill the appropriate requirements of battery electrolyte selection, particular classes of electrolytes, namely organic and inorganic magnesium-aluminum-chloride complexes[12-13] in ethereal solvents, have been developed. Such aforementioned electrolytes exhibit suitable Mg plating/stripping reversibility with a low overpotential and a reasonable anode stability. However, these electrolytes are generally corrosive with rather narrow electrochemical windows.[14] Furthermore, unexpected species in electrolytes can trigger the formation of irreversible passivation layers resistant to the conduction of Mg-ions, thereby limiting the battery voltage induced by the Mg electrode-electrolyte interaction.[14-16]

Consequently, the active search for alternative anode materials, in particular, insertion-type anodes for MIB, is necessary.

Relative to Li- and Na-ion batteries, much less is known about suitable insertion-type anodes and the associated electrochemical reactions for MIBs due to the unique challenge of identifying host materials with appropriate electrochemical capacities that allow for repeated cyclic insertion and deinsertion. Although the ionic radius of Mg (0.72 Å) is smaller than that of Li and Na (0.76 Å and 1.02 Å, respectively), the stronger electrostatic interaction of Mg due to its higher charge/radius ratio slows the insertion kinetics of Mg, which severely limits the possibilities for appropriate insertion-type anode materials for MIBs.[1] Recently, a few materials among group XIV and XV elements are being actively studied as candidate insertion-type anodes for MIBs.[17-19] For example, a thermodynamic analysis of group XIV elements using first principles methods reveals that Ge and Sn have smaller volume expansions (~120% and ~178%, respectively) and diffusion barriers for Mg diffusion (~0.5 and ~0.7 eV, respectively) than Si (~216% and ~1.0 eV).[20-21] This indicates that among this group, Ge and Sn can potentially act as anodes with high gravimetric and volumetric capacity, while still allowing for low operating voltages. Bi and Sn were also experimentally demonstrated to work as high-density anodes with low operating voltages and specific capacities of 384 and 903 mAhg$^{-1}$, respectively.[17-18, 22] While pure Sb fails to show any appreciable capacity and cyclability for Mg,[17] its alloys such as $Bi_{1-x}Sb_x$ and $Sn_{1-x}Sb_x$ also showed a good performance as anode materials for MIBs.[17, 19] Black P was also theoretically reported to be a promising anode for MIBs with optimally low potential of 0.15 V and a high specific capacity of 1730 mAhg$^{-1}$.[23]

While these studies highlight an early promise for group XIV and XV elements as insertion anodes for MIBs, a fundamental and comprehensive understanding of ion insertion/deinsertion reactions (that is, atomic-level mixing/demixing) and the associated microstructural evolution is not yet available. Such understanding is essential for establishing the mechanisms of coupled electrochemical and mechanical behavior, and subsequently for the optimal design of insertion-type anodes. For instance, both experimental and theoretical studies have revealed mechanisms for a two-phase lithiation/delithiation in Si anodes.[24-26] Based on such detailed understanding of insertion/deinsertion mechanisms, core-shell or hollow structures can be designed to effectively alleviate the structural and mechanical degradation of Si-based anodes for Li-ion batteries.[27-28] Moreover, the unique electrochemical properties of Mg (two valence electrons and a smaller ionic radius of 0.72 Å) can be expected to result in unusual insertion/deinsertion behavior of insertion-type anodes not seen for monovalent lithium or sodium ions, which necessitates selection of electrochemically compatible electrolytes for MIBs.

Motivated by the recent experimental demonstrations of group XIV elements as plausible host materials for Mg-ion anodes, and given the wide utilization of these elements as high-capacity anodes for Li-ion[29-30] and Na-ion[31-32] batteries here we investigated the thermodynamics and kinetics of Mg insertion reactions in group XIV X (X = Si, Ge, Sn) anodes. In order to understand the underlying mechanisms of Mg insertion (magnesiation) and the associated microstructural evolution, we employed density functional theory and *ab initio* molecular dynamics methods coupled with ring statistics analysis. We find that Mg insertion can result in the formation of amorphous $Mg_xX$ phases through the formation of weaker Mg-X networks. In addition, Mg ions diffuse much faster in Sn and Ge than in Si, owing to the much weaker Mg-Sn and Mg-Ge bonding conditions. We also conducted thermodynamic studies to investigate the electrochemical performance of magnesiation process in X anodes. We find that the thermodynamically unstable behavior of Mg insertion requires a certain overpotential to prevent Mg aggregation at anode/electrolyte interfaces upon magnesiation. Amorphization serves as an effective and a general approach for decreasing overpotential for crystalline anodes. Finally, amorphous Si and Ge, as well as crystalline Sn have a low average electrode potential in the range of 0.05 – 0.40 V.

## 2. METHODS

The magnesiation behavior of crystalline/amorphous Si and Ge, as well as crystalline Sn anodes, was studied using *ab initio* molecular dynamics (AIMD) simulations. The 2×2×2 supercells of crystalline Si (*c*-Si), Ge (*c*-Ge) (both consisting of 64 atoms) and *β*-Sn (*c*-Sn) (consisting of 48 atoms) were first built. A melt-and-quench procedure was utilized to construct the amorphous phases of *c*-Si and *c*-Ge through AIMD simulations.[33-34] We established the validity of these newly-formed amorphous Si (*a*-Si) and Ge (*a*-Ge) models by calculating radial distribution functions (See **S1** and **Figure S1** in **Supporting Information**). The atomistic models of crystalline and amorphous X (X = Si, Ge, Sn) anodes were then set in direct contact with a liquid Mg reservoir along *z* axis with the number ratio of $N_{Mg}:N_X$ = 1:1, to simulate the dynamic process of Mg insertion at finite temperatures (see **Figure 1** at 0 ps). It has been reported that Mg insertion results in an almost linear volumetric expansion with Mg concentration and reaches up to 216%, 178% and 120% for full magnesiation phase $Mg_2X$ for Si, Ge and Sn, respectively.[20] In order to accommodate such volumetric expansion during Mg-ion insertion, we constrain in-plane (*x-y*) expansion and set the size of supercells along *z* axis equal to the value that leads to the volumetric expansion at Mg concentration $x = 1.0$.

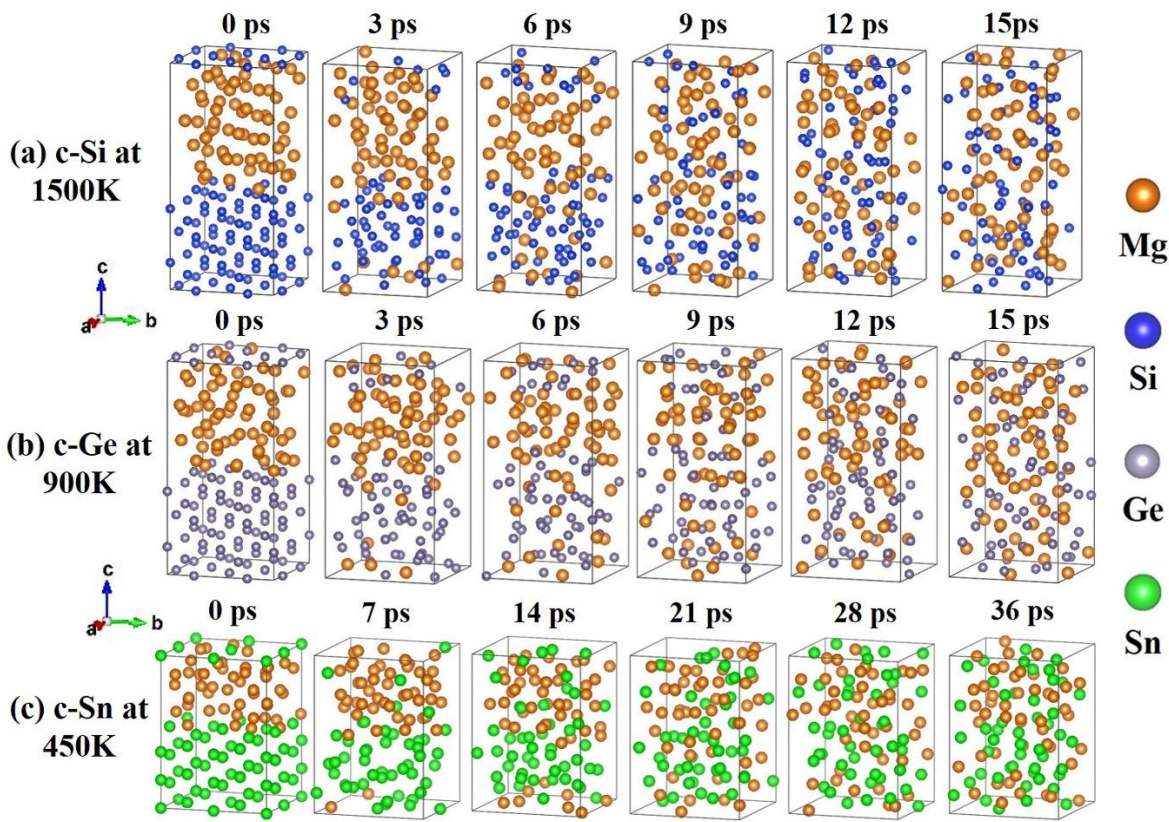

**Figure 1.** Atomistic configurations of crystalline (a) Si, (b) Ge and (c) Sn as a function of reaction time during magnesiation. AIMD simulations were run for crystalline Si, Ge and Sn at the temperatures of 1500 K, 900 K and 450 K, respectively.

AIMD simulations were conducted using plane-wave density functional theory (DFT) methods implemented in Vienna Ab Initio Simulation Package (VASP).[35-36] The projector augmented wave (PAW)[37] pseudopotentials were utilized to describe core and valence electrons. The generalized gradient approximation based on the Perdew-Burke-Ernzerhof (GGA-PBE)[38] function was used to describe electron exchange and correlation. We employed a plane-wave kinetic energy cutoff of 350 eV and 2×2×1 (2×2×2) Monkhorst-Pack[39] *k*-point mesh method for sampling the Brillouin zone and for modelling the Mg-X reaction (amorphization). Convergence tests with respect to the kinetic energy cutoff and *k*-point mesh were performed to validate the accuracy of calculations. To accelerate the Mg insertion process, AIMD simulations were run using a *NVT* ensemble for ~15000 MD time steps of 1 fs for Si and Ge at elevated temperatures in the range of 1200-1500 K, 600-900 K, respectively. To avoid the formation of liquid Mg-Sn system at high temperature, a much lower temperature range of 350-450 K was set for Sn. In such case, a larger time step of 3 fs was selected for Sn to accelerate Mg-Sn reaction.

After Mg-X systems reached their steady state mixed states, the simulations were continued further to estimate ion diffusivities at different temperatures. To realize this, mean square displacements (MSD) of Mg and X (X = Si, Ge, Sn) atoms as a function of simulation time were calculated according to the equation:

$$\text{MSD} = \left\langle \sum_{i=1}^{3} \left| r_i(t) - r_i(0) \right|^2 \right\rangle = \frac{1}{N} \sum_{j=1}^{N} \sum_{i=1}^{3} \left| r_i^j(t) - r_i^j(0) \right|^2 \quad (1)$$

where $N$ is the total atom number; $\langle ... \rangle$ denotes the average value over all atoms; $r_i(0)$ and $r_i(t)$ are $i$-axis positions at simulation time 0 and $t$. The microstructural evolution of Mg-X systems during magnesiation was analysed by evaluating radial distribution functions (RDFs) and changes of the bonding network (ranging from isolated atoms to 6-atom rings). All microstructural analysis was conducted using the R.I.N.G.S. code,[40] which has been utilized previously to explore the lithiation behavior of Si anode[24, 26] and S cathode.[41]

To understand the underlying mechanisms of magnesiation in anodes, it is necessary to conduct the energetic analysis of Mg insertion at different Mg concentration $x$. Both crystalline ($c$-) and amorphous ($a$-) Mg$_x$X phases were considered for calculating the binding energy ($E_b$) and the formation energy ($E_f$) per Mg atom based on the expressions:

$$E_b(x) = \left[ E(\text{Mg}_x\text{X}) - E(\text{X}) - nE(\text{Mg}^{iso}) \right] / n \quad \text{and} \quad (2)$$

$$E_f(x) = \left[ E(\text{Mg}_x\text{X}) - E(\text{X}) - nE(\text{Mg}^{hcp}) \right] / n \;, \quad (3)$$

where $E(Mg_xX)$ and $E(X)$ are the total energies of $Mg_xX$ phases and pure X anodes in their equilibrium crystalline states; $E(Mg^{iso}) = –0.044$ eV is the energy of an isolated Mg atom, $E(Mg^{hcp}) = –1.542$ eV is the energy per atom of Mg in the hcp phase, and $n$ is the total number of Mg atoms in the $Mg_xX$ phase. According to these definitions, a negative value of $E_b$ indicates the chemical binding of Mg-ion to the host anode is thermodynamically favorable, while a negative value of $E_f$ suggests that Mg-X phases are thermodynamically stable against separation into clusters of Mg and X. For the intermediate $c$-$Mg_xX$ phases, since there exist no $c$-$Mg_xX$ structures except for $c$-$Mg_2X$ reported in the literature, we constructed the initial $c$-$Mg_xX$ models with Mg atoms occupying the most stable sites in X anodes[34] (i.e. tetrahedral sites in Si and Ge,[20] and interstitial site enclosed by 5 atoms in Sn[21]). As for $a$-$Mg_xX$ systems, they were obtained from stable $c$-$Mg_xX$ structures by following the same melt-and-quench procedure as outlined above, and then fully optimized until the Hellmann-Feynman forces were less than 0.005 eVÅ$^{-1}$. A plane-wave kinetic energy cutoff of 500 eV and a 3×3×3 $k$-point mesh method were used to estimate $E_b$ and $E_f$ in all cases.

## 3. RESULTS AND DISCUSSION

**Microstructural evolution during Mg insertion.** We first analyze the evolution of atomistic structure and bonding networks of crystalline X (X = Si, Ge, Sn) anodes during Mg insertion. **Figure 1** presents the associated snapshots illustrating the structural evolution. The interfacial Mg atoms gradually react with the bulk crystal of anodes, leading to the breaking of X-X bonds and a volumetric expansion to accommodate the ion insertion. **Figure 2(a-c)** shows the radial distribution functions (RDFs) $g(r)$ of crystalline anodes as a function of time for both X-X and Mg-X pairs. In all cases, the first sharp peak of $g_{X-X}(r)$ pair functions decreases upon Mg insertion, indicating a reduction in the total number of X-X neighbors. Moreover, the second peak of $g_{X-X}(r)$ pair functions gradually disappears, suggesting the formation of amorphous features in the atomistic structures of Mg-X systems. The disappearance of the peaks of $g_{X-X}(r)$ pair functions is also accompanied with a corresponding sharp increase in the peaks of $g_{Mg-X}(r)$ pair functions, indicating the formation of Mg-X bonding networks. The results showing the crystalline-to-amorphous transformation as presented in **Figures 1** and **Figure 2(a-c)** are consistent with recent experiments, which reported TED-EDX characterization of amorphous $Mg_xSn$ formed during magnesiation of Sn anode ($x < 2.0$).[18] It should be noted that generally amorphous structures can transform into crystalline structures upon reaching full ion-insertion capacity, for example, Si and Ge anodes for Li-ion batteries,[42-43] and Ge anode for Na-ion batteries.[44] However, such amorphous-to-crystalline phase transition cannot be captured in our AIMD simulations, since we only consider partial ($x < 2.0$) rather than full magnesiation ($x = 2.0$). Moreover, both intrinsic (length- and time-scale) limitation of AIMD method and unconsidered factors (for example, electric field) can limit the observation of such phase transformation behavior. This can also be seen in the persistence of medium peaks of $g_{X-X}(r)$, which show that there still exist a number of X-X bonding pairs for a period of time after Mg insertion. Nevertheless, similar amorphization phenomena were also reported during the studies of lithiation (for example, in Si anode[24-25, 45]) and sodiation processes (in Si, Ge, Sn anodes) of bulk crystalline anodes.[46] We have also performed similar studies of Mg insertion in *a*-Si and *a*-Ge anodes (see **Figure 2(d-e)**), which show a qualitatively similar behavior to their crystalline counterparts.

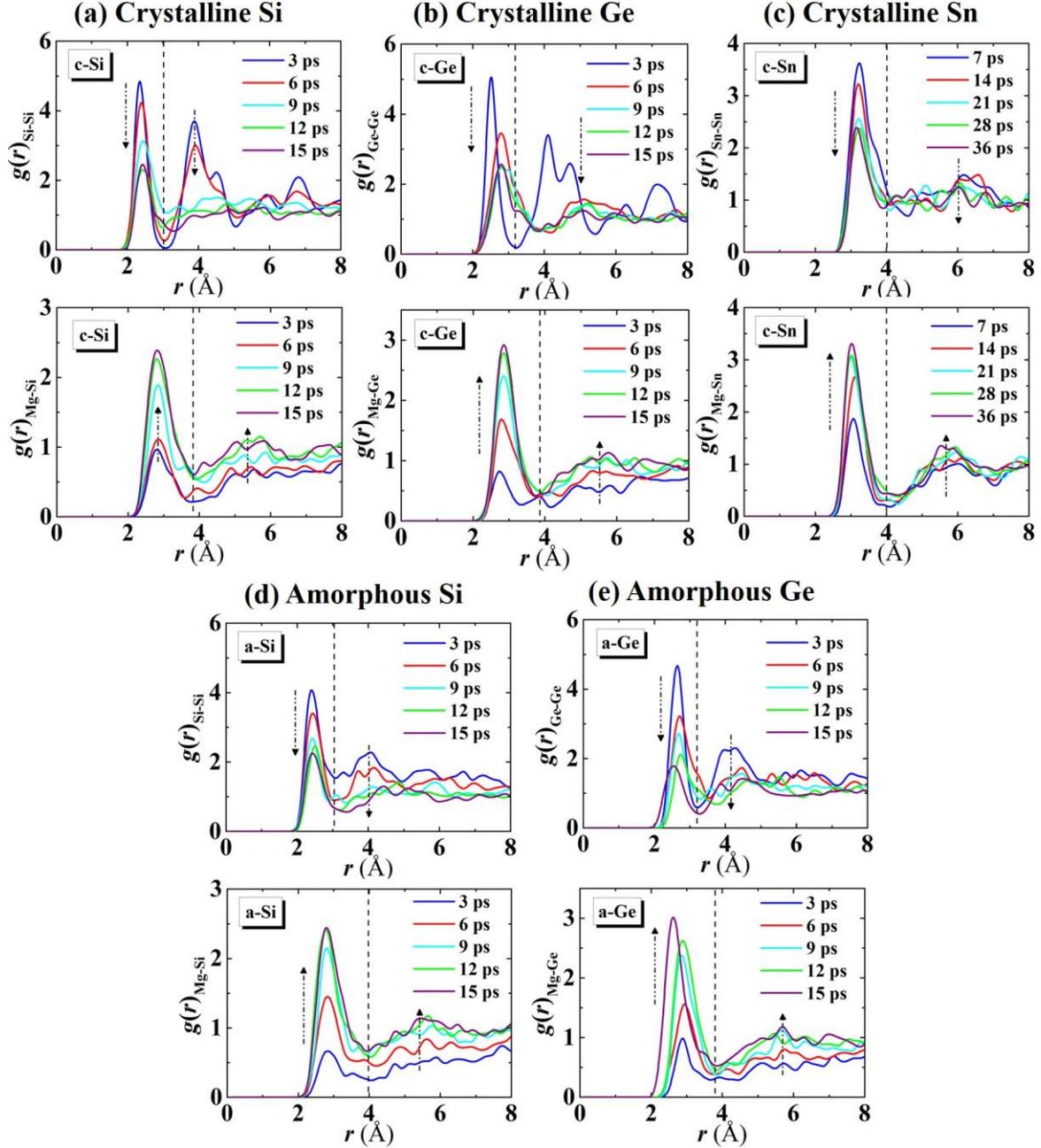

**Figure 2.** Radial distribution functions (RDFs) $g(r)$ of X-X and Mg-X pairs (X = Si, Ge, Sn) in crystalline (a) Si, (b) Ge and (c) Sn at 1500 K, 900 K and 450 K, respectively, and amorphous (d) Si and (e) Ge at 1500 K and 900 K, respectively, as a function of reaction time. The dash arrows represent the variation trend of first and second peaks of $g(r)$.

In order to understand the microstructural evolution of anodes upon Mg insertion, we next evaluate the statistics of covalent-bonded rings and other structural features of X (X = Si, Ge, Sn) upon magnesiation. According to calculated RDFs as shown in **Figure 2**, the cutoff distances for X-X and Mg-X bonds ($r_{X-X}$ and $r_{Mg-X}$) could be determined at the valley right after the first peaks of $g_{X-X}(r)$ and $g_{Mg-X}(r)$. This leads to $r_{Si-Si} = 2.75$ Å, $r_{Mg-Si} = 3.7$ Å, $r_{Ge-Ge} = 2.9$ Å, $r_{Mg-Ge} = 3.8$ Å, $r_{Sn-Sn} = 3.7$ Å, $r_{Mg-Sn} = 3.7$ Å, respectively. These bond cutoff distances were then utilized to determine the bonding networks of X anodes. In general, *c*-Si and *c*-Ge only possess 6-atom rings owing to their hexagonal structures, while *a*-Si and *a*-Ge mainly possess 5- to 7-atom rings resulting from their local distortion. As for *c*-Sn, it possess 4-atom and 5-atom rings owing to its tetragonal structure. **Figure 3** shows that the total number of rings reduces with reaction time as the insertion of Mg ions proceeds. In *c*-Si and *c*-Ge (**Figure 3(a-b)**), specifically, Mg ions first

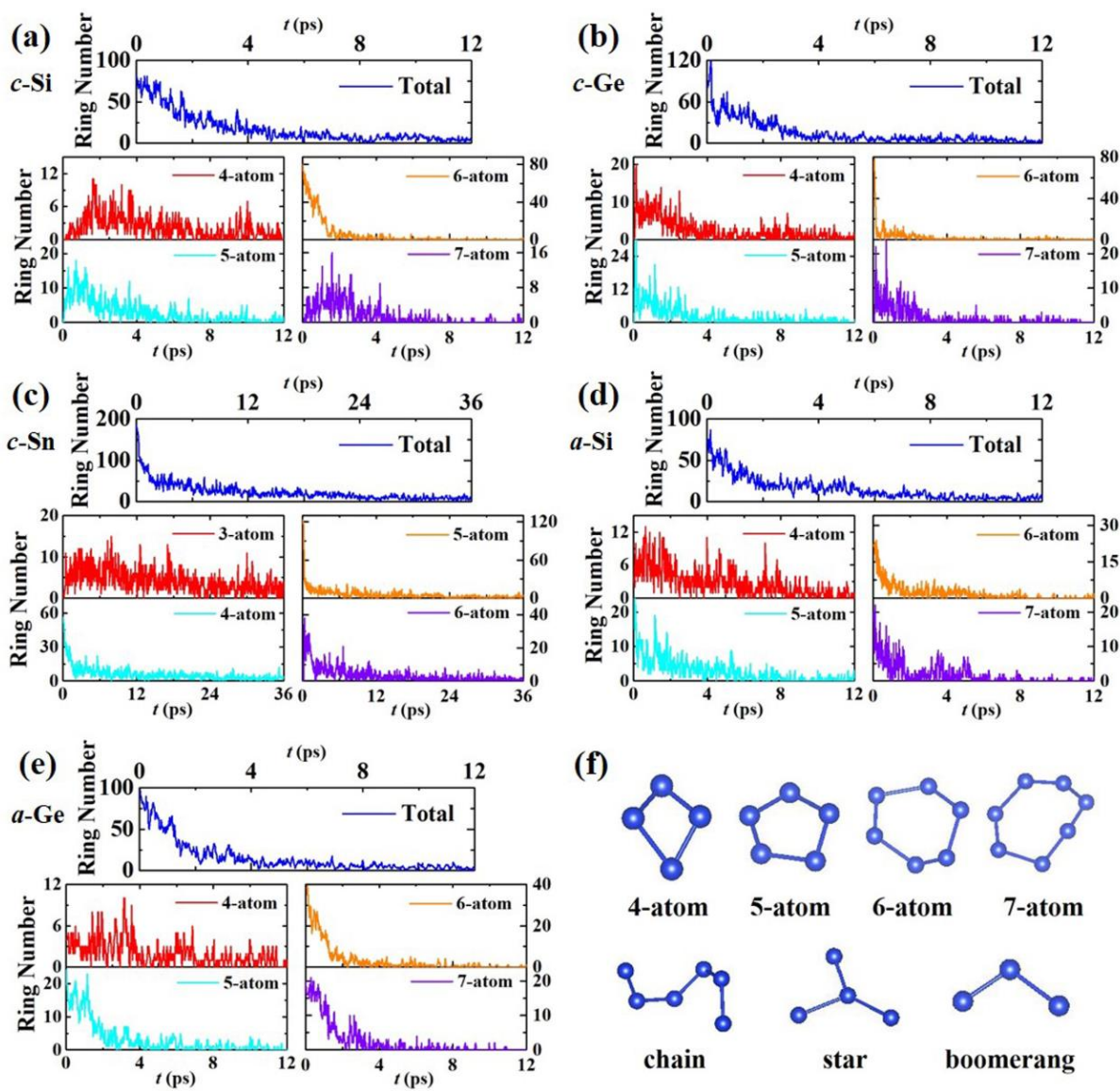

**Figure 3.** Ring statistics of crystalline (a) Si, (b) Ge and (c) Sn at 1500 K, 900 K and 450 K, respectively, and amorphous (d) Si and (e) Ge at 1500 K and 900 K as a function of reaction time. The analysed rings range from 4- to 7-atom rings for Si and Ge, and from 3- to 6-atom rings for Sn. (f) Topological schematic of bonding structures, i.e. rings, chains, stars and boomerangs.

break the X-X bonds of 6-atom rings with a reduced total number, and then generate more 4- to 7-atom rings. After about 3 ps (**Figure 3(a-b)**), the total number of newly-formed rings reduces again with increasing Mg concentration, becoming almost negligible after 8 ps. For *c*-Sn (**Figure 3c**), the majority 5-atom rings are broken to form 3-, 4- and 6-atom rings. A few 3-atom rings still exist even after 30 ps, which can be attributed to the slower Mg-Sn reaction rate at low temperatures, thereby requiring a longer simulation time to form the fully mixed Mg-Sn system. Similar evolution of rings was also found in the cases of *a*-Si and *a*-Ge, as shown in **Figure 3(d-e)**. It was also observed that raising temperature accelerates the reduction of total ring numbers, which can be attributed to the higher Mg-X reaction rates at elevated temperatures. In addition to X-X bonded rings, other X-X bonded structures are also generated during Mg insertion. For instance, X atoms occasionally formed chains, stars and boomerangs during the breaking of rings, but finally dumbbells and isolated atoms dominate and react with Mg ions to form amorphous Mg-X systems. Finally, we found the evolution of the microstructural characteristics such as radial distribution functions and ring statistics shown in **Figures 2** and **3** are qualitatively similar for all temperatures considered in our study. This confirms that elevated temperatures accelerate the formation of the reaction front, as well as the supply of Mg ions required to propagate the front, finally leading to similar microstructural features in the steady state. This observation underscores the utility of AIMD simulations at elevated temperatures for understanding the microstructural evolution in the realistic Mg-ion battery application at room temperatures. Similar methods have also been employed in literature to study reactions in Li- and Na-ion batteries.[24, 47-48]

**Ion diffusion in mixed Mg-X systems.** Ionic diffusivity in microstructural phases of electrode materials is an important aspect for the operating performance of the battery, since it plays a critical role in the charging and discharging dynamics in batteries. In order to determine the equilibrium diffusivities of Mg and X (X = Si, Ge, Sn) species, namely $D_{Mg}$ and $D_X$, we first allowed Mg-X systems to reach their steady state mixed states, namely MgX phases, and then continued the AIMD simulations further for a long time to obtain the average mean square displacements (MSDs) as a function of time at different temperatures. Several earlier studies have employed nudged elastic band method using a single or few ions to estimate diffusivities in battery electrodes, in particular, in intercalation type electrodes.[49] However, for the electrodes with conversion reactions as considered here, AIMD simulations at finite temperatures allow for a direct determination of the diffusivities of ions and host species, as well as an average energy barrier for the diffusion of Mg ions. The AIMD simulation temperatures were selected below the melting points of host anode materials, which ensures that a short-range order is maintained in the amorphous Mg-X systems.[50] The calculated MSDs are presented in **Figures S2-S3,** which show a nearly linear variation with simulation time. Einstein relation $MSD = 6Dt$ was then utilized to calculate the diffusivities of both Mg and X species at different temperatures. For estimating the diffusivities of Mg and X species, it is important to account for the inherent compressive stresses generated due to the insertion of Mg (see **Methods** section). In order to evaluate the intrinsic diffusivities in unstrained conditions, we used the correction[51]

$$D = D_\sigma / \exp(\alpha\sigma/G), \quad (4)$$

where $D$ and $D_\sigma$ are the diffusivities under unstrained and strained conditions, $\alpha$ (typically in the range of 6 – 10) is the coupling parameter, $\sigma$ and $G$ are the residual stress and the shear modulus of the $a$-MgX phase. This relation indicates that the predicted ion diffusivities under compressive (tensile) stress could be significantly underestimated (overestimated) without an appropriate correction. Using the elastic constants predicted by DFT calculations, we used the Voigt-Reuss-Hill approximation[52] to obtain the average shear moduli of $a$-MgX as 21.8, 19.5 and 1.7 GPa for X=Si, Ge, Sn, respectively (see **S2** in **Supporting Information**). After correcting the diffusivities for the inherent stresses at different temperatures, Arrhenius relation[53] $D_T = D_0 \exp(-E_{ba}/k_B T)$ was used to extrapolate the diffusivities at 300 K, $E_{ba}$, $k_B$, $T$, $D_0$ being the energy barrier, the Boltzmann constant, temperature and the prefactor. **Figure 4** shows the intrinsic diffusivities of Mg and X atoms at 300 K for different values of coupling parameters for all anodes considered here. The magnitudes of the residual compressive stresses are listed in **Table S1**. Even though ion migration is driven by potential difference within anodes in realistic batteries, the current high-temperature acceleration method for our diffusivity calculations is still reasonable, and has been successfully employed to explore Li diffusion in Si and Ge anodes.[24, 47]

It is evident that a larger coupling parameter $\alpha$ leads to a larger diffusivity of the same atom species under compressive stress. We select an intermediate value $\alpha = 8$ for the following discussion of the diffusivities. It is evident that Mg atoms (**Figure 4(a)**) have higher diffusivities than X atoms (**Figure 4(b)**) in all anode materials, owing to the weaker Mg-X interactions than X-X interactions in Mg-X material systems. In Si anodes, the diffusivity of Mg is an order of magnitude larger than that of Si atoms. It is interesting to note that the equilibrium diffusivities of Mg and X atoms in crystalline anodes are close to those in amorphous anodes. This is due to a similar amorphous structure of X anodes in their fully mixed states irrespective of the starting structures. Even though the non-equilibrium ion diffusivities at the beginning of Mg-X reaction cannot be predicted by **Equation 1**, the evaluation of energy barriers against Mg diffusion allows to qualitatively understand their diffusivities in the early stages of the reaction. For example, the energy barrier against Mg diffusion in crystalline Si/Ge is larger than that in amorphous Si/Ge.[54] Therefore, we infer that Mg diffusion in crystalline anodes is much slower than that in corresponding amorphous anodes during non-equilibrium charging process. A similar enhancement of Li diffusivity by amorphization has been confirmed in $c$-Si and $a$-Si anodes.[24]

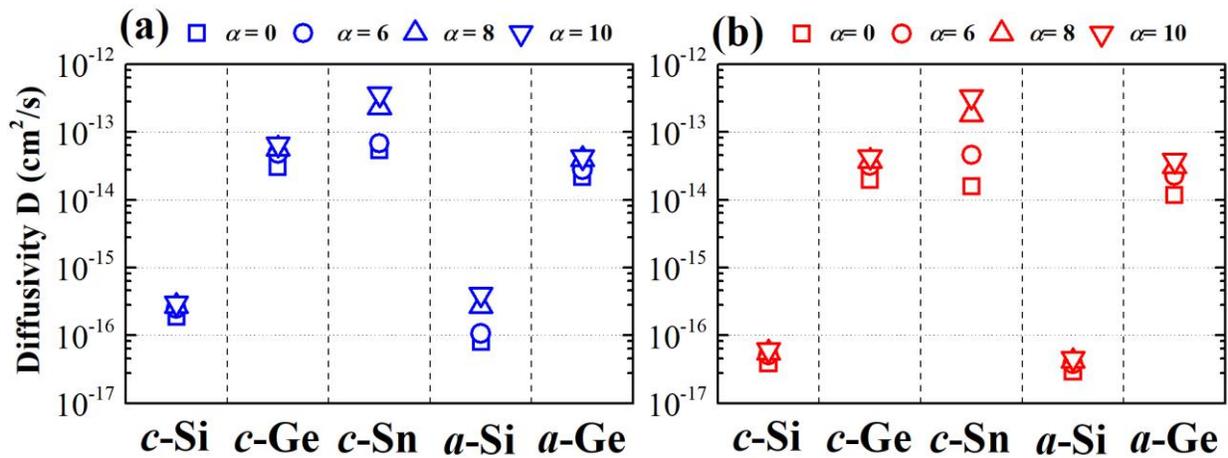

**Figure 4.** Equilibrium ion diffusivities of (a) Mg and (b) X (X = Si, Ge, Sn) species in crystalline (*c*-) and amorphous (*a*-) X anodes at room temperature (300 K). The coupling parameter $\alpha$ in Equation 4 ranges from $\alpha = 0$ to $\alpha = 10$.

Among different anodes considered here, Mg diffusivities follow the trend: $D_{Mg}$ (Si) < $D_{Mg}$ (Ge) < $D_{Mg}$ (Sn). Especially as shown in **Figure 4(a)**, Mg diffusivity in Sn is about 3 orders of magnitude larger than those in Si, and an order of magnitude larger than those in Ge. This also can be understood by the fact that Si possesses the largest energy barrier against Mg diffusion (i.e. 1.42 eV in $c$-Si,[20] while 0.44 eV in $c$-Sn[21]). By fitting the slope of natural logarithm of the Arrhenius relation, we estimate the energy barriers ($E_b$) for Mg and X diffusion in $a$-MgX phases (see **Figure S4** in **Supporting Information**). We find that the values of $E_b$ for average Mg diffusion in $a$-MgX phases are in the range of 0.1 eV–0.3 eV, and are much lower than for single Mg-ion diffusion in crystalline X anodes as reported earlier ($E_b$ in the range of 0.5 eV–1.4 eV[20]). This can be attributed to changes in bonding structures upon Mg insertion—single or few Mg ions are unlikely to cause large scale changes to the stiff X-X bonding structures in crystalline X anodes. However, Mg-X reaction and mixing shown in **Figure 1** breaks the stiffer X-X bonding structures and forms soft, expanded Mg-X bonding structures. In such case, Mg and X species are easier to diffuse in *a*-MgX phases with much lower energy barriers $E_b$.

It is interesting to note that the electrochemical insertion of Mg is characteristically distinct from that of Li in X anodes. For example, the crystalline intermediate phases of Li$_x$X ($0 < x < 4.4$) are thermodynamically stable relative to their amorphous counterparts in contrast to the intermediate phases of Mg$_x$X ($0 < x < 2.0$).[42] However, Mg ion diffusivities in X anodes follow the same trend as those of Li ones ($D_{Li}$ (Si) < $D_{Li}$ (Ge) < $D_{Li}$ (Sn)). This is controlled by the intrinsic material properties such as ionic radii and stiffness of the host Si, Ge and Sn anodes.[55-56] Given the measured Li diffusivities in Si of the order of $10^{-13}$~$10^{-12}$ cm$^2$/s,[57-59] our simulation results show that Mg atoms in Ge and Sn have similar diffusivities (~$10^{-14}$ – $10^{-13}$ cm$^2$/s) to Li in Si within the difference of one order of magnitude, while Mg atoms in Si have much lower diffusivities (~$10^{-17}$ cm$^2$/s) than Li in Si by 3 orders of magnitude. For lithiation process in Si, it is known that diffusion behavior controls the overall lithiation kinetics after sufficient Si has been lithiated by Li-Si reaction.[60] Since similar Mg-X reactions occur in X (X = Si, Ge, Sn) anodes, our kinetic analysis indicates that Ge and Sn with their excellent diffusion behavior and magnesiation kinetics, can be promising candidates as potential insertion-type anodes for MIBs. Finally, it should be noted that for the analysis presented here, we only considered microstructures with stoichiometric compositions without any defects such as vacancies or doping. Recent studies have reported that such defects can further enhance the rates of reaction and ion diffusivities in conversion anodes such as Si and Ge for Li-ion and Na-ion batteries.[61-62] However, an investigation of the role of such defects is beyond the scope of this study.

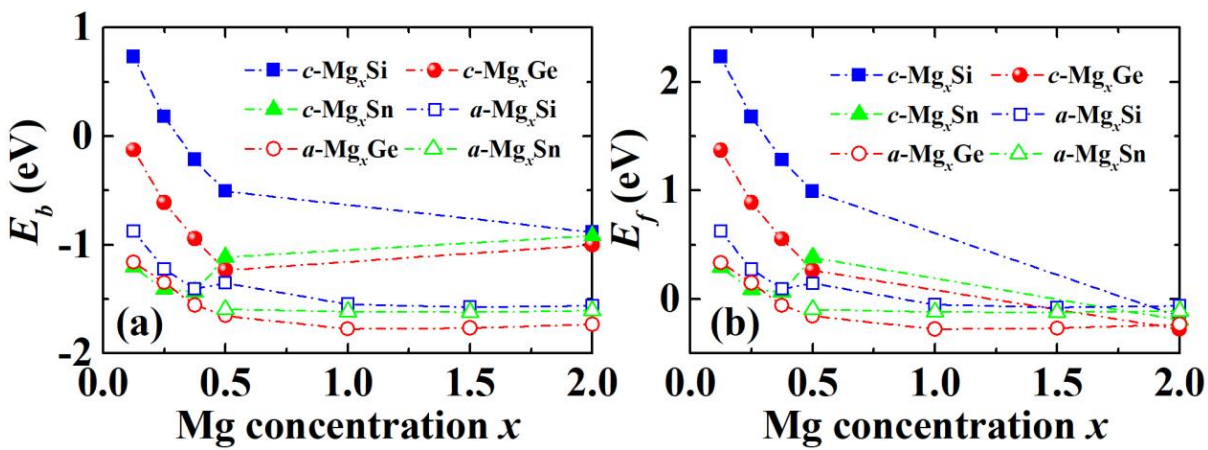

**Figure 5.** Variation of (a) binding energy $E_b$ and (b) formation energy $E_f$ per Mg atom for crystalline (*c*-) and amorphous (*a*-) Mg$_x$X (X = Si, Ge, Sn) phases with Mg concentration $x$.

**Thermodynamic and electrochemical properties of magnesiated anodes.** In order to uncover the underlying mechanisms of magnesiation in X anodes, both binding and formation energies per Mg atom for $Mg_xX$ ($E_b$ and $E_f$) were evaluated using **Equations 2-3**. In general, a negative binding energy $E_b$ indicates that the insertion of a single Mg atom in $Mg_xX$ is thermodynamically favorable and vice versa. On the other hand, a negative formation energy $E_f$ indicates a thermodynamic driving force against the separation into stable forms of both Mg and X clusters and vice versa. The sign of $E_f$ is thus also an indicator to determine the competition between Mg plating at the anode/electrolyte interface and magnesiation in $Mg_xX$ during the charging process. **Figure 5(a-b)** present the calculated values of the binding energies and formation energies. From **Figure 5(a)**, it is evident that the binding of Mg to host anodes is favorable in most cases of $c$- and $a$-$Mg_xX$, except for the $c$-$Mg_xSi$ at lower Mg concentration $x < 0.375$. A large and positive $E_b$ (0.73 eV) for $c$-$Mg_{0.125}Si$ indicates that Mg insertion in the $c$-Si anode is hindered at the beginning of magnesiation process. In contrast, the negative values of $E_b$ for $a$-$Mg_xSi$ reveal that amorphization could facilitate the Mg insertion in Si anode. The absence of Mg insertion in $c$-Si is also consistent with our AIMD simulations of the magnesiation of $c$-Si at lower temperature (T < 900 K), in which case $c$-Si does not react with Mg reservoir since the thermal energy cannot overcome a high activation energy barrier against Mg insertion. Furthermore, it can be seen in **Figure 5(a)** that $E_b$ decreases with Mg concentration, suggesting Mg insertion becomes thermodynamically favorable with subsequent magnesiation. Due to the crystalline structure of pure metallic Sn, $Mg_xSn$ should be still crystalline at low Mg concentration $x$. Thus we only consider the formation of amorphous $Mg_xSn$ above an intermediate Mg concentration $x = 0.5$.

In contrast to the negative binding energies in most cases for both $c$- and $a$-$Mg_xX$, **Figure 5(b)** shows some positive values of formation energy $E_f$ for $c$- and $a$-$Mg_xX$ phases, especially for $c$-$Mg_xSi$ and $c$-$Mg_xGe$ at Mg concentration $x \leq 0.5$. Since the crystalline phases of $Mg_xX$ systems were only reported at $x = 2.0$, we only evaluated the values of $E_f$ for $c$-$Mg_{2.0}X$. The large and positive values of $E_f$ indicate that the inserted Mg prefer to aggregate as Mg clusters, which can be seen as undesirable Mg plating in anode materials, detrimental to the Mg-ion battery performance. **Figure 5(b)** also shows that amorphization can significantly decrease the formation energies for Si and Ge anodes by 1.0–1.6 eV at Mg concentration $x = 0.125$. For $c$-Sn at low Mg concentration ($c$-$Mg_{0.125}Sn$), the formation energy was found to be as low as 0.29 eV. It should be noted that generally, small positive formation energies can be overcome by the application of overpotential to facilitate Mg insertion. In Li-ion batteries, for example, $E_f$ is predicted to be as large as 0.3 eV for c-$Li_{0.015}Si$,[63] but experiments show a small overpotential of 0.1 V could activate the lithiation of $c$-Si during the first cycle of charging process.[64] Therefore, a realistic overpotential can enable the magnesiation of $a$-Si, $a$-Ge and $c$-Sn anodes without Mg plating, while the magnesiation of $c$-Si and $c$-Ge anodes might fail owing to their much higher positive values of $E_f$, which require unrealistically large overpotential.

For Si and Ge anodes at all intermediate Mg concentrations, we find that $E_f$ for $a$-$Mg_xX$ is much lower than that for $c$-$Mg_xX$. It can be concluded that during the magnesiation under large enough overpotential, Mg atoms react with $c$-Si and $c$-Ge anodes and form $a$-Mg$_x$X (X = Si, Ge) phases even at very low concentrations. This mechanism of magnesiation in $c$-Si and $c$-Ge is qualitatively different from the lithiation behavior of $c$-Si, in which both theoretical[34] and experimental[65] studies demonstrate a crystalline-to-amorphous phase transition at low Li concentration ($x \sim 0.3$). Similar to Si and Ge, the formation energies of $a$-Mg$_x$Sn phases are also lower than their crystalline counterparts. The predicted formation of $a$-Mg$_x$X (X = Si, Ge, Sn) in $c$-X anodes at intermediate concentrations is also supported by the RDF analysis of $c$-X anodes upon magnesiation as discussed earlier (see **Figure 2**), in which the sharp decrease of the second peaks of $g_{Mg-X}(r)$ pair functions indicates the quick formation of amorphous Mg$_x$Sn phases in $c$-X anodes. **Figure 5(b)** also shows that at full magnesiation ($x = 2.0$), the $E_f$ for $c$-Mg$_{2.0}$X is slightly lower than that for $a$-Mg$_{2.0}$X, indicating a possible amorphous-to-crystalline transition upon full magnesiation. The formation of $c$-Mg$_{2.0}$Sn has been confirmed by recent experiments using XRD techniques.[18] For comparison, an amorphous-to-crystalline phase transition was reported for $a$-Li$_x$Si when Li concentration reaches a critical value of $x = 3.75$,[42, 66] and which is attributed to the similarity between the electronic structures of $a$-Li$_{3.75}$Si and $c$-Li$_{3.75}$Si.[42] The formation of $c$-Mg$_{2.0}$Si and $c$-Mg$_{2.0}$Ge therefore still requires future experimental verification.

Finally, we also predict the voltage profiles to explore the electrochemical performance of X anodes. Only amorphous phases were considered for all anodes due to their requirement of much lower overpotential as discussed earlier. For every Mg concentration $x$ of $a$-Mg$_x$X (X = Si, Ge, Sn), the electrode potential V($x$) with respect to Mg/Mg$^{2+}$ was calculated as[67]

$$V(x) = -\frac{E(Mg_{x+\Delta x}X) - E(Mg_x X) - \Delta x E(Mg^{hcp})}{ze\Delta x} = \frac{E(Mg_x X) - E(Mg_{x+\Delta x}X)}{ze\Delta x} + \frac{E(Mg^{hcp})}{z} \quad (5)$$

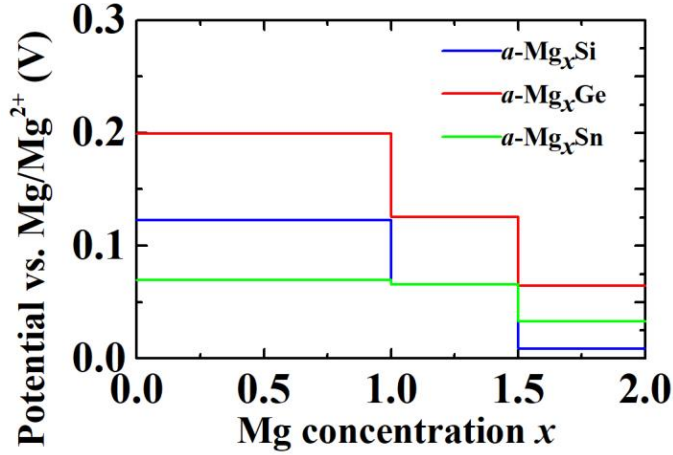

**Figure 6.** Electrode potentials vs. Mg/Mg$^{2+}$ for amorphous (*a*-) Mg$_x$X (X = Si, Ge, Sn) as a function of Mg concentration *x*.

where $E(Mg_xX)$ and $E(Mg_{x+\Delta x}X)$ are the total energies of *a*-Mg$_x$X and *a*-Mg$_{x+\Delta x}$X systems, $\Delta x$ is the increment of Mg concentration, $E(Mg^{hcp}) = -1.542$ eV is the energy per atom of Mg in the hcp phase and $z = 2$ is the nominal charge state for Mg$^{2+}$ ion. **Figure 6** shows the concentration-dependent profile of the electrode potential for *a*-Mg$_x$X. It can be seen that the electrode potential *V* decreases with increasing Mg concentration, and approaches to zero at near full magnesiation. *a*-Mg$_x$Ge has the highest electrode potential at intermediate concentrations, followed by *a*-Mg$_x$Si and *a*-Mg$_x$Sn. In particular, *a*-Mg$_x$Sn exhibits a plateau value of 0.07 eV when $x < 1.5$. Such plateau value predicted by our calculations is lower than the experimentally reported value of 0.15 eV for Sn.[18] Since ideally a good anode should have a low electrode potential and low overpotential, our theoretical results herein suggest that *a*-Si, *a*-Ge and *c*-Sn can be suitable as good anodes for MIB in comparison with *c*-Si and *c*-Ge, owing to their lower overpotential and lower electrode potentials.

In addition to kinetic and thermodynamics of ion insertion, electrical conductivity of electrodes also plays an important role in the overall battery performance. Previous studies of lithiated/sodiated semiconducting anodes have reported that as the concentration of Li/Na ions increases, the electronic band gap of host anode material reduces, and in many cases, a transition from the semiconducting to metallic phase also occurs.[62, 68] Since this behavior is qualitatively similar in alkali and alkaline earth metals, we anticipate Mg insertion can also decrease the band gaps of Mg$_x$X phases and thereby enhance the electrical conductivity. Additional structural design strategies can be also employed to enhance the electrical conductivity of Ge anode for improved battery performance. For example, novel nanostructures (i.e. core-shell, nanoparticle) combined with super conductive additive (i.e. carbon shells, graphene) can be designed to enhance the electrical conductivities of anodes.

## 4. CONCLUSIONS

In summary, *ab initio* molecular dynamics simulations and an energetic analysis based on DFT methods were carried out in order to investigate both the kinetics and thermodynamics of magnesiation (Mg insertion) behavior of group XIV elements as anodes for Mg-ion batteries. We find that Mg insertion in crystalline X anodes can lead to the formation of amorphous Mg$_x$X phases with increasing Mg concentration, accompanied by the breaking of stronger X-X bonding network and the formation of weaker Mg-X bonding network. The Mg diffusivity in Si anodes (~$10^{-17}$ cm$^2$/s) is about two orders of magnitude smaller than Ge and Sn anodes (~$10^{-14} - 10^{-13}$ cm$^2$/s), due to the stronger Mg-Si and Si-Si bonding conditions. Energetic analysis indicated that since the Mg insertion in X anodes at lower Mg concentration is thermodynamically unfavorable, an overpotential would be necessary to facilitate Mg insertion in X anodes and to avoid unexpected Mg plating at the anode/electrolyte interface. In particular, crystalline Si and Ge behave as poor anodes for Mg-ion batteries due to an unrealistically high overpotential, while crystalline Sn can be a suitable anode due to the requirement of much lower overpotential. It was also demonstrated that amorphization serves as an effective approach to decreasing the required overpotential, which would render amorphous Si and Ge as feasible anodes for Mg-ion batteries.

Amorphous Si and Ge, as well as crystalline Sn, have a low average electrode potential in the range of 0.05–0.40 V vs Mg/Mg$^{2+}$. Consequently, the kinetic and electrochemical analysis presented here demonstrate that amorphous Ge and crystalline Sn may possibly serve as promising anodes for Mg-ion batteries. Given the close electrode voltage profiles of magnesiated Ge and Sn to that of Mg metal, promising cathodes (i.e. MnO$_2$ polymorphs and Chevrel Mo$_6$S$_8$)[4-7] and electrolytes (i.e. organic/inorganic magnesium-aluminum-chloride complexes)[12-13] for Mg metal anode as demonstrated in recent experiments may be also utilized with Ge and Sn anodes in Mg-ion batteries. However, the electrochemical full-cell performance as well as anode/electrolyte interfacial behavior needs to be systematically investigated by further experimental and theoretical studies. Analysis of the kinetics and thermodynamics of magnesiation of group XIV elements can also aid in identifying the magnesiation mechanisms in other classes of materials as electrodes for Mg-ion batteries, such as group XV elements (P, As, Sb, Bi) anodes[69] and transition metal dichalcogenides MX$_2$ (M = Ti, Mo, W; X = S, Se, Te) as cathodes.[8-9] Such work can also lead to an optimal design of group XIV anodes via various material synthesis strategies including binary alloying and nano-structural design.[17]

## ASSOCIATED CONTENT

**Supporting Information**

The Supporting Information is available free of charge on the ACS Publications website at DOI. Generation of amorphous Si and Ge; DFT prediction of shear moduli of MgX phases in X (X = Si, Ge, Sn) anodes; supporting figures showing the radial distribution functions of generated amorphous Si and Ge, the predicted mean square displacements of Mg and X atoms in magnesiated X anodes as a function of simulation time at different temperatures, and estimated energy barriers of Mg and X species in crystalline and amorphous X anodes; supporting table exhibiting the average inner stress induced by magnesiation in Mg-X systems (PDF).

## AUTHOR INFORMATION

**Corresponding Author**

*E-mail: mingchao.wang@monash.edu; nikhil.medhekar@monash.edu

**Notes**

The authors declare no competing financial interest.

## ACKNOWLEDGMENT

The authors acknowledge support from the Monash University Cluster, the Australian National Computing Infrastructure (NCI), and the Pawsey Supercomputing Centre for high performance computing. N.V.M. and N.B. gratefully acknowledge the financial support from Australian Research Council's Discovery Project scheme (DP160103661).

# Table of Contents

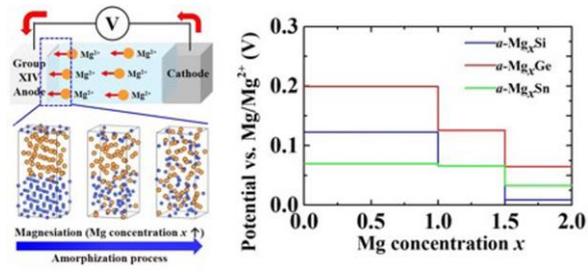